\documentclass[letterpaper, 10 pt, conference]{ieeeconf}  

\IEEEoverridecommandlockouts                              

\usepackage{graphics} 
\graphicspath{ {./images/} }
\usepackage{epsfig} 
\usepackage{cite}
\usepackage{subcaption} 
\usepackage[dvipsnames]{xcolor}
\usepackage{dsfont}   
\usepackage{amsmath} 
\usepackage{amssymb}  
\usepackage{cuted}   
\usepackage{booktabs}
\usepackage{siunitx}
\usepackage{tabularx}
\usepackage{algorithm}
\usepackage{algpseudocode}
\usepackage{multirow}
\usepackage{booktabs,tabularx,subcaption,array}

\title{\LARGE \bf
Modeling and Mixed-Integer Nonlinear MPC of Positive-Negative Pressure Pneumatic Systems}

\author{Yu Mei$^{1}$, Xinyu Zhou$^{1}$ and Xiaobo Tan$^{1}$
\thanks{*This research was supported in part by National Science Foundation awards (ECCS 2024649 and  CNS 2237577).}
\thanks{$^{1}$Yu Mei, Xinyu Zhou and Xiaobo Tan are with the Department of Electrical and Computer Engineering, Michigan State University, East Lansing, MI 48823, USA. 
        Email: {\small meiyu1@msu.edu (YM)}, {\small zhouxi63@msu.edu (XZ)}  {\small xbtan@egr.msu.edu (XT).} }%
}

\begin{document}

\maketitle
\thispagestyle{empty}
\pagestyle{empty}

\begin{abstract}
Positive-negative pressure regulation is critical to soft robotic actuators, enabling large motion ranges and versatile actuation modes. However, it remains challenging due to complex nonlinearities, oscillations, and direction-dependent, piecewise dynamics introduced by affordable pneumatic valves and the bidirectional architecture. We present a model-based control framework that couples a physics-grounded \emph{switched nonlinear} plant model (inflation/deflation modes) with a mixed-integer nonlinear model predictive controller (MI\mbox{-}NMPC). The controller \emph{co-optimizes} mode scheduling and PWM inputs to realize accurate reference tracking while enforcing input constraints and penalizing energy consumption and excessive switching. To make discrete mode decisions tractable, we employ a Combinatorial Integral Approximation that relaxes binary mode variables to continuous surrogates within the valve-scheduling layer. With parameters identified from the physical system, simulations with step and sinusoidal references validate the proposed MI\mbox{-}NMPC, showing a consistently favorable trade-off among accuracy, control effort, and switching, and outperforming conventional PID and NMPC with heuristic mode selection.
\end{abstract}

\section{INTRODUCTION}
Pneumatic systems are widely used in industrial automation \cite{harris2012modelling}, healthcare devices \cite{morales2011pneumatic}, and emerging soft robotics \cite{yap2016high, mosadegh2014pneumatic, mei2024simultaneous, tawk20193d, ambrose2023compact, xie2023soft, mei2023simultaneous, fairchild2023semi, chen2021soft, hashem2020design} owing to their high power-to-weight ratio, high actuation speed, and low cost. In the soft robotics community, soft pneumatic actuators are typically grouped into three categories \cite{zhang2022pneumatic}: positive-pressure actuators \cite{mosadegh2014pneumatic, mei2024simultaneous}, negative-pressure (vacuum-driven) actuators \cite{tawk20193d}, and hybrid-pressure actuators \cite{ambrose2023compact, xie2023soft, mei2023simultaneous, fairchild2023semi, chen2021soft, hashem2020design}. Hybrid-pressure actuators require both positive and negative pressures to enable bidirectional motion or stiffness-tuning mechanisms. It is therefore of interest to develop positive-negative pressure pneumatic systems and associated control schemes that deliver fast, accurate, and energy-efficient regulation under nonlinear dynamics.

Recently, several positive-negative pressure pneumatic systems have been developed \cite{pei2025programmable, park2025modeling, chen2024programmable, zhang2022pneumatic}, and they typically use PID control for the underlying pressure regulation \cite{chen2024programmable, zhang2022pneumatic}. However, because affordable pneumatic components such as on/off valves introduce discontinuous flow rates that excite nonlinear and oscillatory dynamics, fixed-gain PID tuning is often difficult and inadequate \cite{massoud2025enhancing}. To address this challenge, researchers have proposed advanced controllers such as fuzzy piecewise PID \cite{pei2025programmable}, dual-loop PID tuned via evolutionary algorithms \cite{massoud2025enhancing}, and model-based reinforcement learning \cite{park2025modeling}. Among the approaches, model-based controllers are appealing because an explicit plant model allows input and state constraints and the regularization of energy consumption within a unified optimization framework. While the dynamics of on/off solenoid valves and compressible pneumatic flows have been extensively studied \cite{ zhong2021investigation, meng2015system, taghizadeh2009modeling, topccu2006development}, most existing work emphasizes component-level characterization rather than system-level closed-loop pressure regulation. Moreover, dynamic modeling and control of positive-negative pressure pneumatic systems remains limited.

In this work, we develop a mixed-integer nonlinear model predictive controller (MI\mbox{-}NMPC) for positive-negative pressure regulation based on a physics-grounded \emph{switched nonlinear} model. The plant is described by mode-dependent dynamics (inflation/deflation) that capture compressible-flow characteristics, valve dead zones, and direction-dependent flow asymmetries. Based on this switched model, the MI\mbox{-}NMPC \emph{co-optimizes} mode scheduling and PWM inputs to realize the accurate reference tracking while enforcing input constraints and regularizing both energy use and switching behavior. To improve the tractability of discrete mode decisions, we employ a Combinatorial Integral Approximation that relaxes the binary mode variables to continuous surrogates within the MI\mbox{-}NMPC framework for valve scheduling. Comprehensive simulations using parameters identified from the real system show that, compared with conventional PID and NMPC with heuristic mode selection, the proposed approach achieves the most balanced trade-off among tracking accuracy, control effort and reduced switching.

The rest of this paper is organized as follows. Section~II introduces the positive-negative
pressure pneumatic systems, including a brief hardware description, the switched-system modeling of inflation/deflation modes, and a control-affine hybrid formulation. Section~III presents the mixed-integer nonlinear model predictive control (MI\mbox{-}NMPC) design. We propose the mixed-integer optimal control problem and describe the Combinatorial Integral Approximation used for mode scheduling. Section~IV reports the simulation study, detailing the baseline controllers (PID and NMPC), numerical setup, and comparative results. Section~V concludes the paper and outlines directions for future work.

\section{POSITIVE-NEGATIVE PRESSURE PNEUMATIC SYSTEM}

\subsection{System Description}
\begin{figure*}[t]
  \centering
  \includegraphics[width=0.8\textwidth]{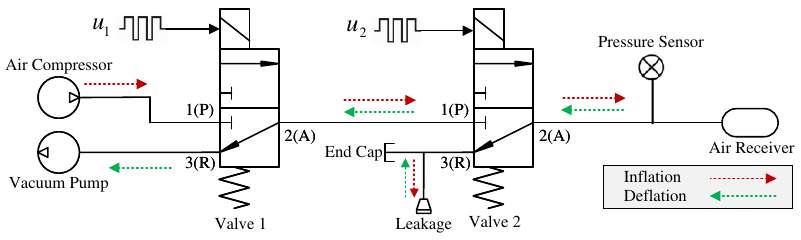}
  \caption{Positive-negative pressure pneumatic system schematic.}
  \label{fig:system}
\end{figure*}

The positive–negative pressure pneumatic system is designed to regulate both positive and negative relative pressure (taking atmospheric pressure as 0 kPa), as illustrated in Fig.~\ref{fig:system}. An air compressor provides positive pressure for inflation, while a vacuum pump supplies negative pressure for deflation. To direct the airflow, two 3/2 solenoid valves are cascaded to control each channel. Each valve operates as a normally closed on/off device, where the switching action is driven by a pulse-width modulation (PWM) signal, denoted as $u_1$ and $u_2$, applied to the coils. When de-energized, airflow is directed from the outlet port (A) to the exhaust port (R), whereas when energized, the inlet port (P) is connected to the outlet port (A). In this configuration, the Valve 1 switches between the constant positive-pressure and negative-pressure sources, with the compressor and vacuum pump connected to its inlet (P) and exhaust (R) ports, respectively. The outlet port (A) of the Valve~1 is then connected to the inlet (P) of the Valve~2, while the exhaust port (R) is blocked with end cap to maintain a vacuum during deflation. Valve~2 controls the airflow delivered to the air receiver, and its pressure is monitored by a pressure sensor in real time. Noting that the leakage at the end cap is also considered to account for practical loss of air. As indicated by the arrows, the red path corresponds to the inflation mode, and the green path corresponds to the deflation mode.

\subsection{Switched System Modeling}
\label{sec:hybrid_model}
The designed positive-negative pressure pneumatic system presents significant challenges for dynamic modeling, since it operates with distinct airflow dynamics in the inflation and deflation modes, resulting in a switched nonlinear system.

\begin{figure}[b]
  \centering
  \includegraphics[scale=0.6]{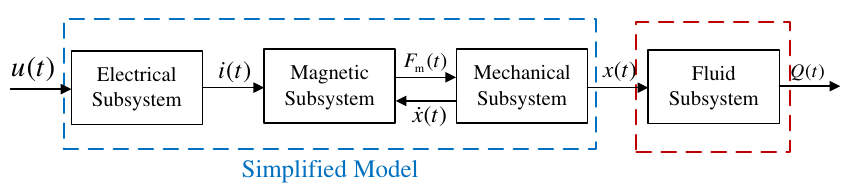}
  \caption{Block diagram of the mathematical model of a solenoid valve subsystems.}
  \label{fig:block}
\end{figure}

The mathematic model of a solenoid valve can be decomposed into four interacting subsystems \cite{meng2015system, topccu2006development}, as shown in Fig.~\ref{fig:block}. The electrical subsystem converts the PWM input voltage $u(t)$ into a coil current $i(t)$. The magnetic subsystem transforms the current into an electromagnetic force $F_m(t)$ acting on the valve spool. The mechanical subsystem then maps this force into spool displacement $x(t)$ and velocity $\dot{x}(t)$, subject to inertia, damping, and spring effects. Finally, the fluid subsystem governs the mass flow rate $Q(t)$ through the valve orifices, which depends on both the spool position and the upstream–downstream pressure differences. Detailed coupling analyses and mathematical derivations of each subsystem are available in \cite{topccu2006development, meng2015system}. In this work, the first three subsystems are simplified into a nonlinear static function $\bar x(u)$, since the rapid transient behavior from the PWM input $u$ to the average spool position $\bar x$ can be neglected in comparison with the dynamics of the other components of the servo-pneumatic system \cite{taghizadeh2009modeling}. In the fluid subsystem, the mass flow rate of an ideal gas through a fixed-area, sharp-edged orifice is modeled by a piecewise function that accounts for the subsonic and choked flow regimes \cite{ISO6358-3, taghizadeh2009modeling}:

\begin{equation}
\label{eq:massflow}
\resizebox{\columnwidth}{!}{$
Q \;=\;
\begin{cases}
P_{\text{up}}\,C_{ud}\,\rho_{\text{ref}}\sqrt{\dfrac{T_{\text{ref}}}{T}}\,
\sqrt{\,1-\left(\dfrac{\dfrac{P_{\text{down}}}{P_{\text{up}}}-b}{1-b}\right)^{\!2}}
\,\bar{x}(u),
& \text{if } b < \dfrac{P_{\text{down}}}{P_{\text{up}}} < 1, \\[6pt]
P_{\text{up}}\,C_{ud}\,\rho_{\text{ref}}\sqrt{\dfrac{T_{\text{ref}}}{T}}\,\bar{x}(u),
& \text{if }\dfrac{P_{\text{down}}}{P_{\text{up}}} \le b \, .
\end{cases}
$}
\end{equation}
All variables and constants in Eq.~\eqref{eq:massflow} are summarized in Table~\ref{tab:pneu_symbols}, together with their physical meanings, units, and values. Among these parameters, the sonic conductance $C$ and the averaged spool function $\bar{x}(u)$ must be experimentally identified. If the averaged spool position $\bar{x}$ is taken as the input, the mass flow rate $Q$ can be expressed as a function of $\bar{x}$, parameterized by $\{P_{\text{up}}, P_{\text{down}}, C_{ud}\}$, and denoted by $Q(\bar{x};P_{\text{up}}, P_{\text{down}}, C_{ud})$. 

\begin{table}[h]
  \caption{Parameters of the pneumatic mass–flow model.}
  \label{tab:pneu_symbols}
  \centering
  \renewcommand{\arraystretch}{1.15}
  \begin{tabularx}{\columnwidth}{@{} l X l l @{}}
    \toprule
    \textbf{Notion} & \textbf{Description} & \textbf{Unit} & \textbf{Value} \\
    \midrule
    $Q$ & Mass flow rate & \si{kg.s^{-1}} & -- \\
    $P_{\text{up}}$ & Upstream absolute pressure & \si{Pa} & -- \\
    $P_{\text{down}}$ & Downstream absolute pressure & \si{Pa} & -- \\
    ${{{P_{{\text{down}}}}} \mathord{\left/
    {\vphantom {{{P_{{\text{down}}}}} {{P_{{\text{up}}}}}}} \right.
    \kern-\nulldelimiterspace} {{P_{{\text{up}}}}}}$ & Pressure ratio & -- & $[0, 1]$ \\
    $b$ & Critical pressure ratio & -- & \num{0.26}$^{\dagger}$ \\
    $C_{ud}$ & Sonic conductance from upstream to downstream & \si{m^{3}.s^{-1}.Pa^{-1}} & $^{\ddagger}$ \\
    $\rho_{\text{ref}}$ & Gas density at reference conditions & \si{kg.m^{-3}} & \num{1.185} \\
    $T_{\text{ref}}$ & Reference temperature & \si{K} & \num{293.15} \\
    $T$ & Gas temperature & \si{K} & \num{293.15} \\
    $\bar{x}$ & Averaged spool position ratio & -- & $[0, 1]$ \\
    $u$ & PWM duty cycle & \% & $[0, 100]$ \\
    $\gamma$ & Heat capacity ratio & -- & \num{1.4} \\
    $R$ & Gas constant for air & \si{J.(kg\,K)^{-1}} & \num{287} \\
    $V$ & Receiver volume & \si{m^{3}} & \num{2.0e-5} \\
    \bottomrule
  \end{tabularx}
  
  \vspace{2pt}
  \footnotesize
  $^{\dagger}$~Value specified in the solenoid valve datasheet (VQ110U, SMC). \\
  $^{\ddagger}$~Parameter needs to be experimentally identified. 
\end{table}

\begin{figure}[b]
  \centering
  \includegraphics[scale=0.6]{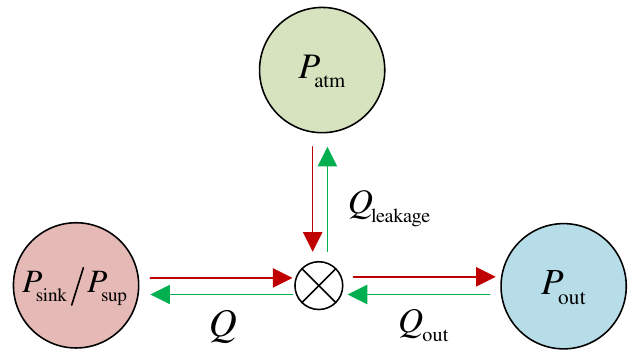}
  \caption{Schematic of the mass flow contributions to the air receiver.}
  \label{fig:flowgraph}
\end{figure}

The effective mass flow rate $Q_{\text{out}}$ delivered to the air receiver is determined by two contributions: the bidirectional mass flow $Q$ through Valve~1, which accounts for both inflation and deflation mode, and the leakage flow $Q_{\text{leakage}}$ through Valve~2 to the atmosphere, as illustrated in Fig.~\ref{fig:flowgraph}. Depending on the receiver pressure $P_{\text{out}}$ relative to the supply $P_{\text{sup}}$, sink $P_{\text{sink}}$, and atmosphere $P_{\text{atm}}$, there exist four different operating cases:
\begin{subequations}\label{eq:qout_cases}
\begin{enumerate}

  \item Inflation process with $P_{\text{atm}} < P_{\text{out}} < P_{\text{sup}}$:
  \begin{equation}
    \resizebox{0.9\columnwidth}{!}{$
    \begin{aligned}
      Q_{\text{out}} &= Q - Q_{\text{leakage}} \\
                       &= Q(\bar{x};\, P_{\text{sup}},\, P_{\text{out}},\, C_{so}) 
                        \;-\; Q(1-\bar{x};\, P_{\text{out}},\, P_{\text{atm}},\, C_{oa}),
    \end{aligned}
    $}
  \end{equation}

  \item Inflation process with $P_{\text{out}} < P_{\text{atm}} < P_{\text{sup}}$:
  \begin{equation}
    \resizebox{0.9\columnwidth}{!}{$
    \begin{aligned}
      Q_{\text{out}} &= Q + Q_{\text{leakage}} \\
                       &= Q(\bar{x};\, P_{\text{sup}},\, P_{\text{out}},\, C_{so})
                        \;+\; Q(1-\bar{x};\, P_{\text{atm}},\, P_{\text{out}},\, C_{ao}),
    \end{aligned}
    $}
  \end{equation}

  \item Deflation process with $P_{\text{sink}} < P_{\text{atm}} < P_{\text{out}}$:
  \begin{equation}
    \resizebox{0.9\columnwidth}{!}{$
    \begin{aligned}
      Q_{\text{out}} &= -\,Q - Q_{\text{leakage}} \\
                       &= -\,Q(\bar{x};\, P_{\text{out}},\, P_{\text{sink}},\, C_{os})
                         \;-\; Q(1-\bar{x};\, P_{\text{out}},\, P_{\text{atm}},\, C_{oa}),
    \end{aligned}
    $}
  \end{equation}

  \item Deflation process with $P_{\text{sink}} < P_{\text{out}} < P_{\text{atm}}$:
  \begin{equation}
    \resizebox{0.9\columnwidth}{!}{$
    \begin{aligned}
      Q_{\text{out}} &= -\,Q + Q_{\text{leakage}} \\
                       &= -\,Q(\bar{x};\, P_{\text{out}},\, P_{\text{sink}},\, C_{os})
                         \;+\; Q(1-\bar{x};\, P_{\text{atm}},\, P_{\text{out}},\, C_{ao}),
    \end{aligned}
    $}
  \end{equation}

\end{enumerate}
\end{subequations}
Here, $P_{\text{sup}}$, $P_{\text{sink}}$, and $P_{\text{atm}}$ denote the positive supply pressure, the negative sink pressure, and the atmospheric pressure, respectively. The leakage flow $Q_{\text{leakage}}$ is modeled by the complementary path $(1-\bar{x})$, consistent with Eq.~\eqref{eq:massflow}. The conductances $C_{so}$, $C_{os}$, $C_{oa}$, and $C_{ao}$ represent the flow paths supply--outlet, outlet--sink, outlet--atmosphere, and atmosphere--outlet, respectively, and are identified experimentally. After obtaining the effective mass flow rate $Q_{\text{out}}$, the pressure dynamics of the air receiver can be derived from the ideal gas law \cite{topccu2006development}:
\begin{equation}\label{eq:idealgas}
{\dot P_{{\text{out}}}} = \gamma \frac{{RT}}{V}Q_{\text{out}},
\end{equation}
where the variables are the constants of the ideal gas law, and their definitions and values are summarized in Table~\ref{tab:pneu_symbols}.

\subsection{Control-affine Hybrid Formulation}
After deriving the piecewise mass–flow dynamics of the orifice valve and the mode-dependent effective outlet flow, the compact control–affine hybrid model can be reformulated by combining Eqs.~\eqref{eq:massflow}, \eqref{eq:qout_cases}, and \eqref{eq:idealgas}:
\begin{subequations}\label{eq:controlform}
\begin{equation}
\dot P_{\text{out}} = f(P_{\text{out}}) + g_m(P_{\text{out}})\,\bar{x}(u), 
\end{equation}
where
\begin{equation}
f(P_{\text{out}}) = \frac{\gamma RT}{V}\,\big[ A_{ao}(P_{\text{out}}) - A_{oa}(P_{\text{out}}) \big],
\end{equation}
\begin{equation}
\resizebox{\columnwidth}{!}{$
\begin{aligned}
g_m(P_{\text{out}}) &= \frac{\gamma RT}{V}\,\Big(
    m\,[A_{so}(P_{\text{out}}) - A_{ao}(P_{\text{out}}) + A_{oa}(P_{\text{out}})] \\
&\quad + (1-m)\,[-A_{os}(P_{\text{out}}) - A_{ao}(P_{\text{out}}) + A_{oa}(P_{\text{out}})]
\Big),
\end{aligned}
$}
\end{equation}
\end{subequations}
where ${P_{{\text{out}}}}$ is the system state and $\bar x(u)$ is the control input. $m$ is a binary mode variable indicating the operating mode:
\begin{equation}
m =
\begin{cases}
1, & \text{if inflation}, \\[4pt]
0, & \text{if deflation}.
\end{cases}
\end{equation}
The functions $A_{so}(\cdot)$, $A_{os}(\cdot)$, $A_{oa}(\cdot)$, and $A_{ao}(\cdot)$ represent the effective flow functions of the four corresponding flow branches:  
\begingroup
\setlength{\jot}{6pt}
\begin{subequations}\label{eq:Afunctions}
\begin{align}
A_{so}(P) &= P_{\text{sup}}\,C_{so}\,\rho_{\text{ref}}
             \sqrt{\tfrac{T_{\text{ref}}}{T}}\,
             \phi\!\left(\tfrac{P}{P_{\text{sup}}}\right), \\ 
A_{os}(P) &= P\,C_{os}\,\rho_{\text{ref}}
             \sqrt{\tfrac{T_{\text{ref}}}{T}}\,
             \phi\!\left(\tfrac{P_{\text{sink}}}{P}\right), \\ 
A_{oa}(P) &= P\,C_{oa}\,\rho_{\text{ref}}
             \sqrt{\tfrac{T_{\text{ref}}}{T}}\,
             \phi\!\left(\tfrac{P_{\text{atm}}}{P}\right), \\ 
A_{ao}(P) &= P_{\text{atm}}\,C_{ao}\,\rho_{\text{ref}}
             \sqrt{\tfrac{T_{\text{ref}}}{T}}\,
             \phi\!\left(\tfrac{P}{P_{\text{atm}}}\right),
\end{align}
\end{subequations}
\endgroup
where $\phi(r)$ is the shape factor defined by the piecewise function
\begin{equation}
\phi(r) =
\begin{cases}
1, & r \leq b, \\[6pt]
\sqrt{\,1 - \left(\dfrac{r-b}{1-b}\right)^{2}}, & b < r < 1, \\[6pt]
0, & r \geq 1 ,
\end{cases}
\end{equation}
with $b$ denoting the critical pressure ratio. 

\section{MIXED-INTEGER NONLINEAR MODEL PREDICTIVE CONTROL}
In this section, we develop a fast mixed-integer nonlinear model predictive controller for the proposed positive-negative pressure pneumatic system, which generates optimal control inputs that track the reference trajectory while considering input constraints, energy consumption and switching penalties.

\subsection{Mixed-integer Optimal Control Problem}
Considering the developed control–affine mode-switching dynamics in Eq.~\eqref{eq:controlform}, the mixed-integer optimal control problem (MIOCP) is formulated by embedding the integer mode variable $m$ into the classical predictive control framework, where both continuous control inputs $\bar{x}(u)$ and discrete switching decisions are jointly optimized over the prediction horizon:
\begingroup
\setlength{\jot}{2pt}%
\setlength{\abovedisplayskip}{6pt}%
\setlength{\belowdisplayskip}{6pt}%
\begin{subequations}\label{eq:miocp}
\begin{align}
\min_{u_I(\cdot),\,u_D(\cdot),\,m(\cdot)}\;\;
& \int_{t_0}^{t_f} \Big[
   q_e\big(P(t)-P_{\text{ref}}(t)\big)^2 \nonumber \\[-2pt]
 +\, & r_I\big(m(t)\,u_I(t)^2 + (1-m(t))\,u_D(t)^2\big)
   \Big]\,dt
\label{eq:miocp:a}
\end{align}

\vspace{-10pt}
\begin{alignat}{2}
\text{s.t.}\;& \dot P = f(P) + g_m(P)\bar{x}(u),
&&\label{eq:miocp:b}\\
& P(t_0)=P_0,
&&\label{eq:miocp:c}\\
& m(t)\in\{0,1\},
&&\label{eq:miocp:d}\\
& u_I^{\min}\le u_I(t)\le u_I^{\max},\quad
  u_D^{\min}\le u_D(t)\le u_D^{\max}.
&&\label{eq:miocp:e}
\end{alignat}
\end{subequations}
\endgroup
where $P(t)$ denotes the outlet pressure state (i.e., $P(t) \equiv P_{\text{out}}(t)$), $P_{\text{ref}}(t)$ is the reference trajectory, $u_I$ and $u_D$ are the inflation and deflation inputs, $u_I^{\min},u_I^{\max},u_D^{\min},u_D^{\max}$ denote the actuator bounds, and $m$ encodes the mode selection.

\subsection{Mixed-Integer NMPC with Combinatorial Integral Approximation}

When implemented with time discretization, the continuous-time MIOCP in Eq.~\eqref{eq:miocp} gives rise to a finite-dimensional Mixed-Integer Nonlinear Program (MINLP). However, MINLPs are generally intractable for real-time control due to their combinatorial complexity, strong nonlinearities, and often nonconvex structures \cite{belotti2013mixed, burger2018algorithm, kirches2013mixed}. To address this challenge, the Combinatorial Integral Approximation technique is adopted. In particular, the binary mode variable $m(t)\in\{0,1\}$ is relaxed to a continuous surrogate $\omega(t)\in[0,1]$. This relaxation decomposes the MINLP into a Nonlinear Program (NLP). The relaxed NLP can be solved efficiently to provide a lower bound to the original MIOCP, while subsequent rounding schemes are employed to recover feasible integer solutions suitable for real-time implementation. After this outer convexification, the MIOCP in Eq.~\eqref{eq:miocp} can be rewritten as:

\begingroup
\setlength{\jot}{2pt} 
\begin{subequations}\label{eq:miocp_disc}
\begin{align}
\min_{\{u_{I,k},\,u_{D,k},\,\omega_k\}_{k=0}^{N-1}}\;\;
&\sum_{k=0}^{N-1} \Big[
   q_e\big(P_k - P_{\text{ref},k}\big)^2 \nonumber\\[-2pt]
 +&\, r_I\big(\omega_k\,u_{I,k}^2 + (1-\omega_k)\,u_{D,k}^2\big) \nonumber\\[-2pt]
 +&\, \lambda_{\text{bin}}\,\omega_k\!\left(1-\omega_k\right)
\Big],
\label{eq:miocp_disc:a}
\end{align}
\vspace{-20pt}
\begin{alignat}{2}
\text{s.t.}\;& P_{k+1} = \Phi_{\mathrm{RK4}}\!\big(P_k,\,u_{I,k},\,u_{D,k},\,\omega_k;\,\delta t\big),
&&\label{eq:miocp_disc:b}\\
& P_0 = P(t_k),
&&\label{eq:miocp_disc:c}\\
& 0 \le \omega_k \le 1,
&&\label{eq:miocp_disc:d}\\
& u_I^{\min}\le u_{I,k}\le u_I^{\max},\quad
  u_D^{\min}\le u_{D,k}\le u_D^{\max},
&&\label{eq:miocp_disc:e}
\end{alignat}
\end{subequations}
\endgroup
where $P_k$ denotes the outlet pressure state at step $k$, 
$P_{\text{ref},k}$ is the reference trajectory, 
and $u_{I,k}$, $u_{D,k}$ are the inflation and deflation PWM inputs, respectively. Eq.~\eqref{eq:miocp_disc:b} represents the fourth-order Runge--Kutta integration of the outer-convexified continuous dynamics with step size $\delta t$:
\begin{equation} \label{outconvexified}
\dot P \;=\; f(P) \;+\; 
\Big[ \,\omega\,g_{m=1}(P) + (1-\omega)\,g_{m=0}(P) \,\Big]\,\bar{x}(u),
\end{equation}
with $g_{m=1}(P)$ and $g_{m=0}(P)$ denoting the mode-dependent dynamics in inflation and deflation, respectively. In addition, the term $\lambda_{\text{bin}}\,\omega_k(1-\omega_k)$ is included in the objective as a relaxation penalty. This term discourages intermediate values of $\omega_k$ and promotes binary mode selections, thereby reducing unnecessary mode switching.

After obtaining the relaxed optimal solution $\{\omega_k^{\mathrm{rel}},\,u_{I,k}^{\mathrm{rel}},\,u_{D,k}^{\mathrm{rel}}\}$ from the NLP, the binary mode decisions are recovered using standard rounding: 
\begin{equation} \label{SR}
m_k^\ast \;=\;
\begin{cases}
1, & \omega_k^{\mathrm{rel}} \geq 0.5, \\[4pt]
0, & \text{otherwise}.
\end{cases}
\end{equation}
Subsequently, the recovered $m_k^\ast$ is substituted for $\omega_k$ in Eq.~\eqref{eq:miocp_disc:b}, and the NLP is re-solved to obtain the finalized solution $\{m_k^\ast,\,u_{I,k}^\ast,\,u_{D,k}^\ast\}$. The first control action $\{m_0^\ast,\,u_{I,0}^\ast,\,u_{D,0}^\ast\}$ is then applied at the current step. The overall procedure is summarized in Algorithm~\ref{alg:CIA_MINMPC}.

\renewcommand{\algorithmicrequire}{\textbf{Inputs:}}
\renewcommand{\algorithmicensure}{\textbf{Outputs:}}
\begin{algorithm}[h]
\caption{\footnotesize \textbf{MINMPC with Combinatorial Integral Approximation}}
\label{alg:CIA_MINMPC}
\begin{algorithmic}[1]
\Require Current state $P(t_k)$; reference $\{P_{\text{ref},k}\}_{k=0}^{N-1}$; horizon $N$; step $\delta t$; bounds $\big(u_I^{\min},u_I^{\max},u_D^{\min},u_D^{\max}\big)$.
\Ensure $\{m_0^\ast,\,u_{I,0}^\ast,\,u_{D,0}^\ast\}$
\State \textbf{Preformulate system constraint:} discretized dynamics \emph{Eq.~\eqref{eq:miocp_disc:b}} from the outer–convexified model \emph{Eq.~\eqref{outconvexified}}.
\State \textbf{Build Relaxed MINLP} with decision vars $\{P_k,u_{I,k},u_{D,k},\omega_k\}_{k=0}^{N-1}$, objective \emph{Eq.~\eqref{eq:miocp_disc:a}}, constraints \emph{Eq.~\eqref{eq:miocp_disc:b}–\eqref{eq:miocp_disc:e}}.
\State \textbf{Solve Relaxed MINLP} $\rightarrow (P^{\mathrm{rel}},u_I^{\mathrm{rel}},u_D^{\mathrm{rel}},\omega^{\mathrm{rel}})$.
\State \textbf{Recover Binary Mode:} determine $m_k^\ast$ from $\omega_k^{\mathrm{rel}}$ according to Eq.~\eqref{SR}
\State \textbf{Fixed-Mode NLP:} replace $\omega_k$ by $m_k^\ast$ in \emph{Eq.~\eqref{eq:miocp_disc:a}–\eqref{eq:miocp_disc:e}} and solve $\Rightarrow (P^\ast,u_I^\ast,u_D^\ast)$.
\State \textbf{Apply} $m_0^\ast$ and $u_0^\ast \!=\! m_0^\ast u_{I,0}^\ast + (1-m_0^\ast)u_{D,0}^\ast$ at the current time $t_k$.
\end{algorithmic}
\end{algorithm}

\section{SIMULATION}
In this section, the feasibility and effectiveness of the proposed MI-NMPC scheme are demonstrated through numerical simulations. Performance is evaluated on step and sinusoidal pressure tracking tasks, and compared against baseline controllers, including a heuristic mode-selection PID and a nonlinear MPC.

\subsection{Baseline Controllers: PID and NMPC}
For benchmarking purposes, two baseline controllers are proposed. For both baseline controllers, the current control mode $m_0$ is heuristically determined from the sign of the current tracking error $e_k=P_{\text{ref},k}-P_k$ at time $t_k$:
\begin{equation} \label{eq:mode_selection}
m_0 \;=\;
\begin{cases}
1, & e_k \geq 0 \quad \text{(inflation mode)}, \\[4pt]
0, & e_k < 0 \quad \text{(deflaion mode)}.
\end{cases}
\end{equation}

For the PID baseline, two separate PID controllers are assigned and individually tuned for the inflation and deflation modes. Let $K_p^{(I)},K_i^{(I)},K_d^{(I)}$ and
$K_p^{(D)},K_i^{(D)},K_d^{(D)}\vphantom{\bigg( \sum\nolimits^{X}_{X} \bigg)}$
denote the PID gains for inflation and deflation, respectively, and let $\delta t$ be the sampling time. Define the integral state $z_k$ (mode–specific) as $z_{k+1}=z_k + e_k\delta t$. The raw PID commands for the two modes are:
\begin{subequations}\label{eq:pid_raw}
{\footnotesize
\begin{align}
u_{I,k}^{\text{raw}} &= K_p^{(I)} e_k + K_i^{(I)} z_k + K_d^{(I)} \frac{e_k - e_{k-1}}{\delta t}, \quad (m=1), \\
u_{D,k}^{\text{raw}} &= K_p^{(D)} e_k + K_i^{(D)} z_k + K_d^{(D)} \frac{e_k - e_{k-1}}{\delta t}, \quad (m=0).
\end{align}
}
\end{subequations}
After performing deadzone-aware linear normalization to the admissible PWM ranges, one ge
{\small
\begin{subequations}\label{eq:PID_mapping}
\begin{align}
u_{I,k} &= u_I^{\min} + \frac{u_I^{\max}-u_I^{\min}}{100}\;u_{I,k}^{\text{raw}}, 
\label{eq:PID_mapping:a}\\
u_{D,k} &= u_D^{\min} + \frac{u_D^{\max}-u_D^{\min}}{100}\;u_{D,k}^{\text{raw}}, 
\label{eq:PID_mapping:b}
\end{align}
\end{subequations}
}

For the NMPC baseline, the system model corresponding to the currently selected mode is employed in the prediction, and the optimization is solved accordingly at each step. In particular, this corresponds to building and solving a fixed-mode NLP at every sampling instant, following the procedure of Step~5--6 in Algorithm~\ref{alg:CIA_MINMPC}, where the mode $m_k$ is same as $m_0$ determined using Eq.~\eqref{eq:mode_selection}.

\subsection{Numerical Setup and Results}
Although the validation is performed in simulation, the system dynamics are modeled to closely approximate the real hardware. All simulations are conducted on the identified hybrid pneumatic system model with the main physical parameters summarized in Table~\ref{tab:params}. The simulations are implemented in Python, with the hybrid pneumatic system dynamics integrated using the forward Euler method. Both MI-NMPC and NMPC formulations are implemented using CasADi~3.7.0. All programs are executed on a desktop computer equipped with an AMD~Ryzen~9~9900X CPU and 32~GB of RAM.

\begin{table}[b]
\centering
\caption{Simulation System Parameters}
\label{tab:params}
\begin{tabular}{ll}
\toprule
\textbf{Parameter} & \textbf{Value} \\
\midrule
Supply pressure $P_{\mathrm{sup}}$ & $300\,\text{kPa}$ \\
Sink pressure $P_{\mathrm{sink}}$ & $10\,\text{kPa}$ \\
Atmospheric pressure $P_{\mathrm{atm}}$ & $100\,\text{kPa}$ \\
Sonic conductances & 
$C_{so}=2.64{\times}10^{-10} \si{m^{3}.s^{-1}.Pa^{-1}}$, \\
& $C_{os}=3.44{\times}10^{-10} \si{m^{3}.s^{-1}.Pa^{-1}}$, \\
& $C_{oa}=6.94{\times}10^{-12} \si{m^{3}.s^{-1}.Pa^{-1}}$, \\
& $C_{ao}=4.52{\times}10^{-12} \si{m^{3}.s^{-1}.Pa^{-1}}$ \\
Valve dead-zones & $u_I^{\min}=20\%$, $u_D^{\min}=25\%$ \\
\bottomrule
\end{tabular}
\end{table}

For the proposed MI-NMPC, the prediction horizon is set to $N=10$ with a step size of $\delta t=0.02\,\text{s}$. The cost weights are chosen as $(q_e,r_I,\lambda_{\text{bin}})=(1,10^{-2},10^{2})$. For the PID baseline, two sets of controllers are designed to address different objectives: steady-state accuracy and transient tracking performance. The first is a gentle PID controller tuned for small steady-state error and smooth response, with gains $(K_p^{(I)},K_i^{(I)},K_d^{(I)};\;K_p^{(D)},K_i^{(D)},K_d^{(D)}) = (0.002,0.0008,0.000;\;0.010,0.001,0.000)$. The second is an aggressive PID controller intended for fast trajectory tracking, with gains $(K_p^{(I)},K_i^{(I)},K_d^{(I)};\;K_p^{(D)},K_i^{(D)},K_d^{(D)}) = (0.004,0.000,0.001;\;0.020,0.000,0.001)$. For the NMPC baseline, the horizon and step size are set identically to the MI-NMPC, and the cost weights are $(q_e,r_I)=(1,3\times10^{-4})$. It is emphasized that the weight parameters for all four controllers (MI-NMPC, NMPC, GENTLE-PID, and AGGR-PID) are carefully tuned to yield their best achievable tracking performance. We evaluate the proposed MI-NMPC together with three baseline controllers under both step and sinusoidal pressure tracking tasks. 
For the step response, the reference trajectory consists of relative pressure plateaus 
$(0,\,40,\,80,\,120,\,80,\,40,\,0,\,-40,\,-80,\,-40,\,0)\,\text{kPa}$, each step holding for $2.0\,\text{s}$. This challenges the controller with large transitions in both inflation and deflation. 
For the sinusoidal case, the reference is defined as 
$P_{\text{ref}}(t) = 40 \sin(2\pi t)\,\text{kPa}$, 
which evaluates continuous tracking performance.

\begin{figure}[t]
  \centering
  \includegraphics[width=1\linewidth]{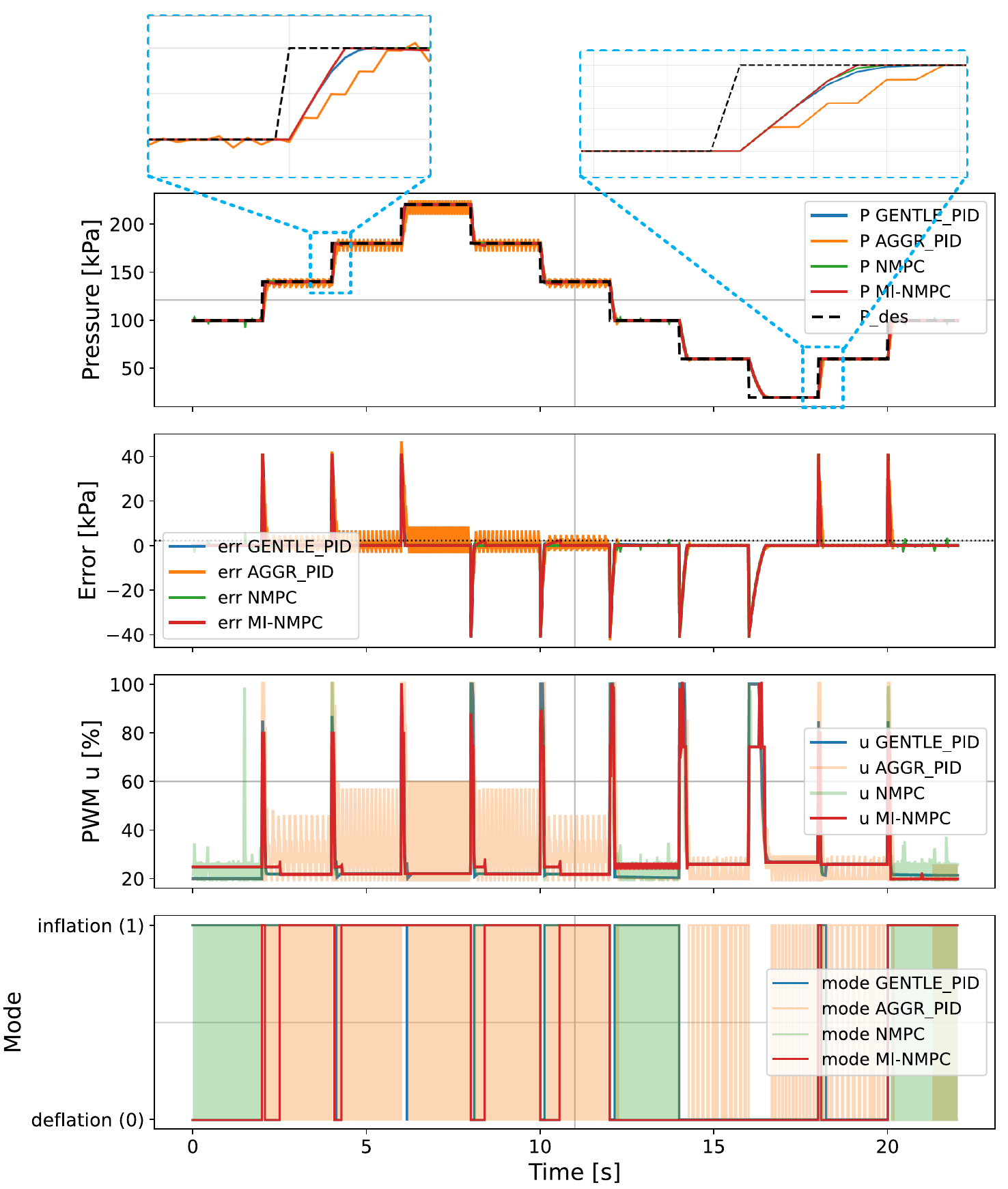}
  \caption{Simulation results of tracking step reference with zoomed-in snapshot. For each case, the subplots from top to bottom show: output pressure, tracking error, control input (PWM), and mode sequence over time.}
  \label{fig:results_step}
\end{figure} 

\begin{figure}[t]
  \centering
  \includegraphics[width=1\linewidth]{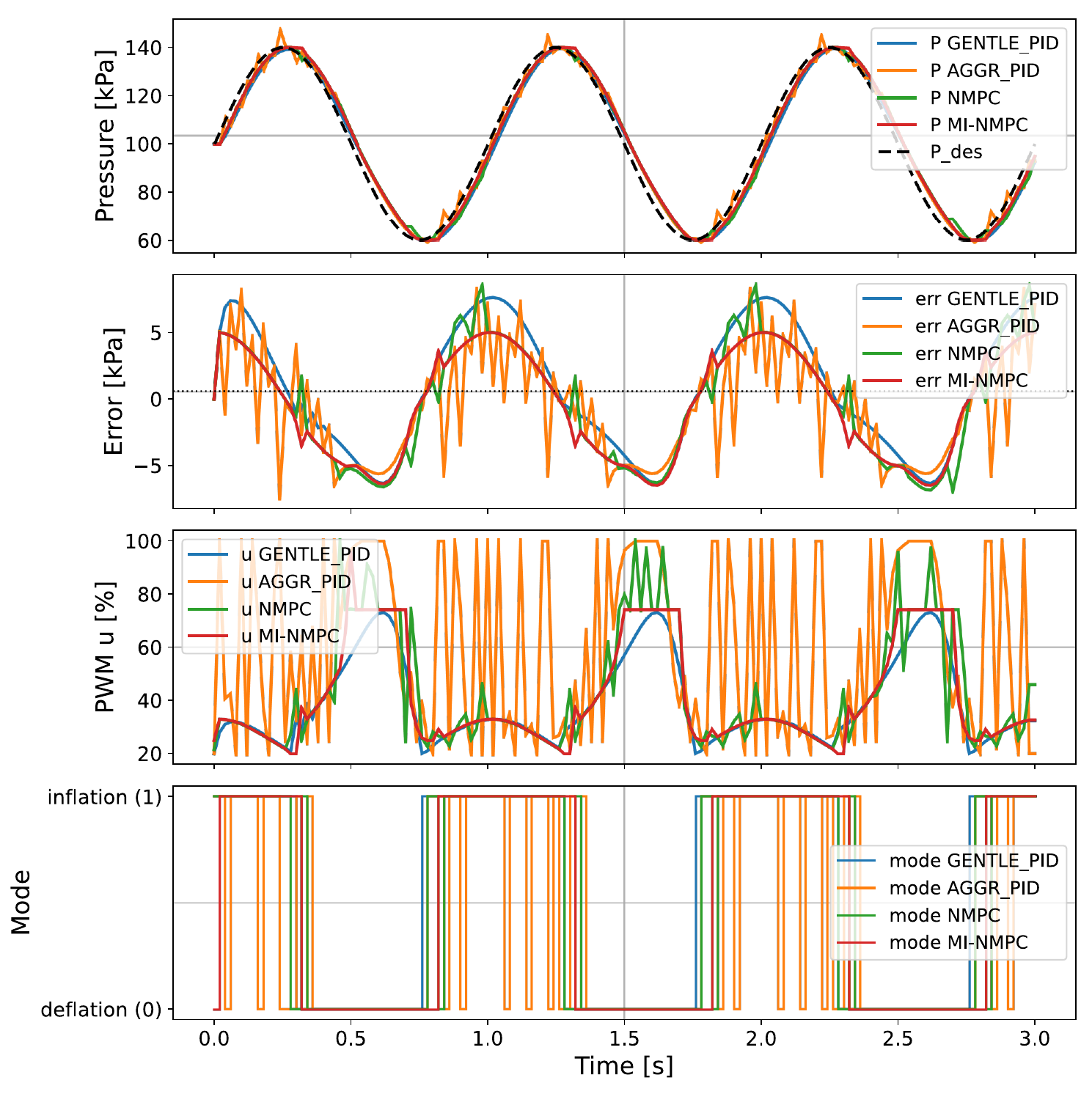}
  \caption{Simulation results of tracking the sinusoidal reference.}
  \label{fig:results_sine}
\end{figure} 

The corresponding simulation results are presented in Fig.~\ref{fig:results_step} and Fig.~\ref{fig:results_sine}. All four controllers are able to track the reference trajectories, but their tracking accuracy, mode-switching behavior, energy consumption and computation time differ significantly. A quantitative comparison is summarized in Table~\ref{tab:results}. In the step reference tracking, the gentle PID achieves an average error of 1.73 kPa, whereas both NMPC and MI-NMPC further reduce the error to 1.40 kPa and 1.42 kPa, respectively. As illustrated in the zoomed-in snapshots of Fig.~\ref{fig:results_step}, the model-based controllers exhibit superior transient behavior. The simple mode-selection NMPC achieves slightly lower tracking error and input energy, but at the cost of excessive mode switching. In the sinusoidal reference tracking with a fast-varying $1\,\text{Hz}$ trajectory, the gentle PID struggles to follow the reference, resulting in large tracking errors of 4.11 kPa. The aggressive PID improves tracking performance but introduces frequent mode switching (46) and more input effect (183.2). By contrast, both NMPC and MI-NMPC achieve comparable tracking accuracy, with MI-NMPC offering smaller maximum tracking error and reduced switching. 

\begin{figure}[t]
  \centering
  \includegraphics[scale=0.35]{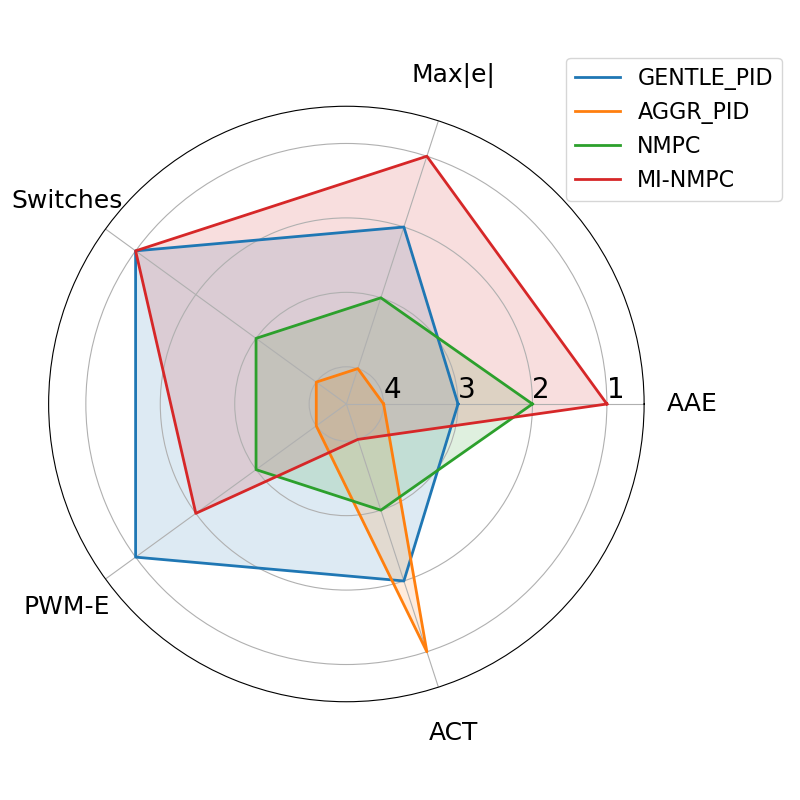}
  \caption{Radar-chart comparison of different controllers across key metrics. \emph{Farther from the center is better}.}
  \label{fig:radar}
\end{figure}

In summary, the proposed MI-NMPC achieves the most balanced trade-off among tracking accuracy, control effort, and mode switching. It delivers consistent tracking with smoother mode transitions, while the baseline controllers either suffer from larger steady-state errors or incur excessive switching when the reference changes rapidly. However, this performance comes at the cost of significantly higher computational demand. The proposed MI-NMPC requires on the order of hundreds of milliseconds to compute a single step, whereas the baseline PID and NMPC controllers operate within a few milliseconds or less. This computational burden could potentially be reduced by advanced implementation techniques reported in other works \cite{quirynen2024real, gratzer2024two, gros2020reinforcement}. To visualize the overall trade-offs, Fig.~\ref{fig:radar} shows a radar chart that averages the step and sinusoidal results across key metrics, plotted as ranks with the best value on the outer ring, and the large covered area highlights MI-NMPC’s consistently favorable performance. 

\begin{table}[t]
\centering
\begingroup
\scriptsize
\setlength{\tabcolsep}{3pt}
\caption{Performance comparison of controllers under step and sinusoidal references.}
\label{tab:results}

\subcaption*{(a) Step reference}
\begin{tabularx}{\columnwidth}{l *{5}{>{\centering\arraybackslash}X}}
\toprule
Controller & AAE [kPa] & Max$|e|$ [kPa] & Switches & PWM-E [\%$\cdot$s] & ACT [ms] \\
\midrule
GENTLE-PID     & 1.73 & 40.61 & 14  & 583.0 & 0.00205 \\
AGGR-PID       & 2.48 & 45.78 & 432 & 672.9 & 0.00169 \\
NMPC           & 1.40 & 40.11 & 259 & 593.5 & 12.69023 \\
MI-NMPC        & 1.42 & 40.00 & 13  & 597.6 & 122.63131 \\
\bottomrule
\end{tabularx}

\vspace{0.6em}

\subcaption*{(b) Sinusoidal reference}
\begin{tabularx}{\columnwidth}{l *{5}{>{\centering\arraybackslash}X}}
\toprule
Controller & AAE [kPa] & Max$|e|$ [kPa] & Switches & PWM-E [\%$\cdot$s] & ACT [ms] \\
\midrule
GENTLE-PID     & 4.11 & 7.64 & 6  & 118.1 & 0.00211 \\
AGGR-PID       & 3.54 & 8.25 & 46 & 183.2 & 0.00198 \\
NMPC           & 3.89 & 8.55 & 18 & 130.5 & 24.27057 \\
MI-NMPC        & 3.65 & 6.50 & 7  & 125.1 & 300.62712 \\
\bottomrule
\end{tabularx}
\endgroup

\vspace{0.25em}
\begin{minipage}{\columnwidth}
\scriptsize
\textit{Notes:} AAE = Average absolute error; Max$|e|$ = Maximum absolute error; PWM-E = Integrated PWM energy; ACT = Average computation time per step. \emph{Lower is better.}
\end{minipage}
\end{table}

\section{CONCLUSION}
This paper presented a model-based framework for positive-negative pressure regulation that couples a physics-grounded switched nonlinear model with a mixed-integer nonlinear MPC. By co-optimizing mode scheduling and PWM inputs while handling input constraints, the controller achieved consistent tracking with reduced switching, outperforming conventional PID and NMPC with heuristic mode selection in simulation using real-system parameters. 

Future work will focus on real-time implementation and solver acceleration to close the computation gap, and hardware validation on multi-channel soft actuators with robustness testing under model uncertainty and disturbance. Additional directions include formal stability analysis and integrating lightweight learning-based residuals for improved model fidelity.



\clearpage            
\bibliographystyle{IEEEtran}
\bibliography{acc}

\begin{thebibliography}{10}
\providecommand{\url}[1]{#1}
\csname url@samestyle\endcsname
\providecommand{\newblock}{\relax}
\providecommand{\bibinfo}[2]{#2}
\providecommand{\BIBentrySTDinterwordspacing}{\spaceskip=0pt\relax}
\providecommand{\BIBentryALTinterwordstretchfactor}{4}
\providecommand{\BIBentryALTinterwordspacing}{\spaceskip=\fontdimen2\font plus
\BIBentryALTinterwordstretchfactor\fontdimen3\font minus \fontdimen4\font\relax}
\providecommand{\BIBforeignlanguage}[2]{{%
\expandafter\ifx\csname l@#1\endcsname\relax
\typeout{** WARNING: IEEEtran.bst: No hyphenation pattern has been}%
\typeout{** loaded for the language `#1'. Using the pattern for}%
\typeout{** the default language instead.}%
\else
\language=\csname l@#1\endcsname
\fi
#2}}
\providecommand{\BIBdecl}{\relax}
\BIBdecl

\bibitem{harris2012modelling}
P.~G. Harris, G.~E. O’Donnell, and T.~Whelan, ``Modelling and identification of industrial pneumatic drive system,'' \emph{The International Journal of Advanced Manufacturing Technology}, vol.~58, no.~9, pp. 1075--1086, 2012.

\bibitem{morales2011pneumatic}
R.~Morales, F.~J. Badesa, N.~Garc{\'\i}a-Aracil, J.~M. Sabater, and C.~P{\'e}rez-Vidal, ``Pneumatic robotic systems for upper limb rehabilitation,'' \emph{Medical \& Biological Engineering \& Computing}, vol.~49, no.~10, pp. 1145--1156, 2011.

\bibitem{yap2016high}
H.~K. Yap, H.~Y. Ng, and C.-H. Yeow, ``High-force soft printable pneumatics for soft robotic applications,'' \emph{Soft Robotics}, vol.~3, no.~3, pp. 144--158, 2016.

\bibitem{mosadegh2014pneumatic}
B.~Mosadegh, P.~Polygerinos, C.~Keplinger, S.~Wennstedt, R.~F. Shepherd, U.~Gupta, J.~Shim, K.~Bertoldi, C.~J. Walsh, and G.~M. Whitesides, ``Pneumatic networks for soft robotics that actuate rapidly,'' \emph{Advanced Functional Materials}, vol.~24, no.~15, pp. 2163--2170, 2014.

\bibitem{mei2024simultaneous}
Y.~Mei, L.~Peng, H.~Shi, X.~Qi, Y.~Deng, V.~Srivastava, and X.~Tan, ``Simultaneous shape reconstruction and force estimation of soft bending actuators using distributed inductive curvature sensors,'' \emph{IEEE/ASME Transactions on Mechatronics}, vol.~29, no.~4, pp. 2849--2857, 2024.

\bibitem{tawk20193d}
C.~Tawk, G.~M. Spinks, M.~in~het Panhuis, and G.~Alici, ``{3D} printable linear soft vacuum actuators: their modeling, performance quantification and application in soft robotic systems,'' \emph{IEEE/ASME Transactions on Mechatronics}, vol.~24, no.~5, pp. 2118--2129, 2019.

\bibitem{ambrose2023compact}
J.~W. Ambrose, N.~Z.~R. Chiang, D.~S.~Y. Cheah, and C.-H. Yeow, ``Compact multilayer extension actuators for reconfigurable soft robots,'' \emph{Soft Robotics}, vol.~10, no.~2, pp. 301--313, 2023.

\bibitem{xie2023soft}
Z.~Xie, M.~Mohanakrishnan, P.~Wang, J.~Liu, W.~Xin, Z.~Tang, L.~Wen, and C.~Laschi, ``Soft robotic arm with extensible stiffening layer,'' \emph{IEEE Robotics and Automation Letters}, vol.~8, no.~6, pp. 3597--3604, 2023.

\bibitem{mei2023simultaneous}
Y.~Mei, P.~Fairchild, V.~Srivastava, C.~Cao, and X.~Tan, ``Simultaneous motion and stiffness control for soft pneumatic manipulators based on a lagrangian-based dynamic model,'' in \emph{2023 American Control Conference (ACC)}, 2023, pp. 145--152.

\bibitem{fairchild2023semi}
P.~Fairchild, N.~Shephard, Y.~Mei, and X.~Tan, ``Semi-physical modeling of soft pneumatic actuators with stiffness tuning,'' \emph{ASME Letters in Dynamic Systems and Control}, vol.~3, no.~4, p. 041006, 2023.

\bibitem{chen2021soft}
F.~Chen, Y.~Miao, G.~Gu, and X.~Zhu, ``Soft twisting pneumatic actuators enabled by freeform surface design,'' \emph{IEEE Robotics and Automation Letters}, vol.~6, no.~3, pp. 5253--5260, 2021.

\bibitem{hashem2020design}
R.~Hashem, M.~Stommel, L.~K. Cheng, and W.~Xu, ``Design and characterization of a bellows-driven soft pneumatic actuator,'' \emph{IEEE/ASME Transactions on Mechatronics}, vol.~26, no.~5, pp. 2327--2338, 2020.

\bibitem{zhang2022pneumatic}
B.~Zhang, J.~Chen, X.~Ma, Y.~Wu, X.~Zhang, and H.~Liao, ``Pneumatic system capable of supplying programmable pressure states for soft robots,'' \emph{Soft Robotics}, vol.~9, no.~5, pp. 1001--1013, 2022.

\bibitem{pei2025programmable}
X.~Pei, R.~Shi, L.~Wang, S.~Liu, Z.~Wu, and Z.~Dai, ``A programmable pneumatic system with novel improved controller enabling adhesion and desorption operations of soft adhesive robots,'' \emph{IEEE Transactions on Automation Science and Engineering}, vol.~22, pp. 19\,986--19\,999, 2025.

\bibitem{park2025modeling}
S.~H. Park, M.~Doh, C.~Park, A.~T. Luong, H.~R. Choi, J.~C. Koo, H.~Rodrigue, and H.~Moon, ``Modeling and reinforcement learning-based control of simultaneous positive and negative pressure generation in pneumatic systems,'' \emph{IEEE Robotics and Automation Letters}, vol.~10, no.~7, pp. 7516--7523, 2025.

\bibitem{chen2024programmable}
P.~Chen, Q.~Ding, Y.~Liu, Z.~Deng, and J.~Huang, ``Programmable pressure control in pneumatic soft robots with 2-way 2-state solenoid valves,'' \emph{IEEE Robotics and Automation Letters}, vol.~9, no.~7, pp. 6448--6455, 2024.

\bibitem{massoud2025enhancing}
M.~M. Massoud, P.~H. Alves, and J.~Libby, ``Enhancing dual-loop pressure control in pneumatic soft robotics with a comparison of evolutionary algorithms for {PID} \& {FOPID} controller tuning,'' \emph{IEEE Robotics and Automation Letters}, vol.~10, no.~6, pp. 6119--6126, 2025.

\bibitem{zhong2021investigation}
Q.~Zhong, X.~Wang, H.~Zhou, G.~Xie, H.~Hong, Y.~Li, B.~Chen, and H.~Yang, ``Investigation into the adjustable dynamic characteristic of the high-speed/valve with an advanced pulsewidth modulation control algorithm,'' \emph{IEEE/ASME Transactions on Mechatronics}, vol.~27, no.~5, pp. 3784--3797, 2021.

\bibitem{meng2015system}
F.~Meng, H.~Zhang, D.~Cao, and H.~Chen, ``System modeling, coupling analysis, and experimental validation of a proportional pressure valve with pulsewidth modulation control,'' \emph{IEEE/ASME Transactions on Mechatronics}, vol.~21, no.~3, pp. 1742--1753, 2015.

\bibitem{taghizadeh2009modeling}
M.~Taghizadeh, A.~Ghaffari, and F.~Najafi, ``Modeling and identification of a solenoid valve for pwm control applications,'' \emph{Comptes Rendus Mecanique}, vol. 337, no.~3, pp. 131--140, 2009.

\bibitem{topccu2006development}
E.~E. Top{\c{c}}u, {\.I}.~Y{\"u}ksel, and Z.~Kam{\i}{\c{s}}, ``Development of electro-pneumatic fast switching valve and investigation of its characteristics,'' \emph{Mechatronics}, vol.~16, no.~6, pp. 365--378, 2006.

\bibitem{ISO6358-3}
\emph{ISO 6358-3:2014, Pneumatic fluid power -- Determination of flow-rate characteristics of components using compressible fluids -- Part 3: Method for measuring steady-state flow-rate characteristics of pneumatic valves}, International Organization for Standardization Std., 2014.

\bibitem{belotti2013mixed}
P.~Belotti, C.~Kirches, S.~Leyffer, J.~Linderoth, J.~Luedtke, and A.~Mahajan, ``Mixed-integer nonlinear optimization,'' \emph{Acta Numerica}, vol.~22, pp. 1--131, 2013.

\bibitem{burger2018algorithm}
A.~B{\"u}rger, C.~Zeile, A.~Altmann-Dieses, S.~Sager, and M.~Diehl, ``An algorithm for mixed-integer optimal control of solar thermal climate systems with mpc-capable runtime,'' in \emph{2018 European Control Conference (ECC)}.\hskip 1em plus 0.5em minus 0.4em\relax IEEE, 2018, pp. 1379--1385.

\bibitem{kirches2013mixed}
C.~Kirches, H.~G. Bock, J.~P. Schl{\"o}der, and S.~Sager, ``Mixed-integer nmpc for predictive cruise control of heavy-duty trucks,'' in \emph{2013 European control conference (ECC)}.\hskip 1em plus 0.5em minus 0.4em\relax IEEE, 2013, pp. 4118--4123.

\bibitem{quirynen2024real}
R.~Quirynen, S.~Safaoui, and S.~Di~Cairano, ``Real-time mixed-integer quadratic programming for vehicle decision-making and motion planning,'' \emph{IEEE Transactions on Control Systems Technology}, vol.~33, no.~1, pp. 77--91, 2025.

\bibitem{gratzer2024two}
A.~L. Gratzer, M.~M. Broger, A.~Schirrer, and S.~Jakubek, ``Two-layer mpc architecture for efficient mixed-integer-informed obstacle avoidance in real-time,'' \emph{IEEE Transactions on Intelligent Transportation Systems}, vol.~25, no.~10, pp. 13\,767--13\,784, 2024.

\bibitem{gros2020reinforcement}
S.~Gros and M.~Zanon, ``Reinforcement learning for mixed-integer problems based on mpc,'' \emph{IFAC-PapersOnLine}, vol.~53, no.~2, pp. 5219--5224, 2020.

\end{thebibliography}







\end{document}